\documentclass{iaus}
\usepackage{graphicx}

\title[Central mass accumulation in nuclear spirals] 
{Central mass accumulation in nuclear spirals} 

\author[W. Maciejewski]   
{Witold Maciejewski}

\affiliation{Astrophysics Research Institute, Liverpool John Moores University,
Twelve Quays House, Egerton Wharf, Birkenhead CH41 1LD, UK; email: wxm@astro.livjm.ac.uk}

\pubyear{2007}
\volume{245}  
\jname{Proceedings Title IAU Symposium}
\editors{A.C. Editor, B.D. Editor \& C.E. Editor, eds.}
\begin{document}

\maketitle

\begin{abstract}
In central regions of non-axisymmetric galaxies high-resolution hydrodynamical 
simulations indicate spiral shocks, which are capable of transporting gas 
inwards. The efficiency of transport is lower at 
smaller radii, therefore instead of all gas dropping onto the galactic centre, 
a roughly uniform distribution of high-density gas develops in the gaseous
nuclear spiral downstream from the shock, and the shear in gas is very low 
there. These are excellent conditions for star formation. This mechanism is 
likely to contribute to the process of (pseudo-) bulge formation. 
\end{abstract}

\firstsection 
\section{Introduction}
Galaxies with non-axisymmetric mass distribution induce efficient gas inflow,
which can span throughout most of the galaxy, leading to the accumulation of 
gas in the galaxy centre, on scales comparable to the resolution limits
of observations or computation. In some cases, high-resolution data indicate 
presence of nuclear rings in the innermost few hundred parsecs of a galaxy, 
where gas density is high and star formation occurs. Kormendy \& Kennicutt 
(2004) postulated that stars created there eventually give rise to 
pseudobulges. Although formation of nuclear rings can be studied in detail 
with grid-based hydrodynamical modelling within the Eulerian scheme 
(e.g. Piner et al 1995), 
evolution of galaxies is most often studied within the Lagrangian scheme
(e.g. SPH methods), which appears more flexible. However, these studies usually
do not resolve the central gas concentration resulting from inflow (e.g. Di
Matteo et al. 2007), usually called the central blob. SPH studies that
resolve nuclear rings in some cases, in other cases still produce a central 
blob (Patsis \& Athanassoula 2000). On the other hand, 
high-resolution Eulerian models indicate that aside for nuclear rings,
nuclear spirals can form in centres of galaxies (Englmaier \& Shlosman
2000, Maciejewski 2000, Maciejewski et al. 2002). Recently, most detailed
studies within the Lagrangian scheme (Ann \& Thakur 2005) demonstrated that
the central blob can be resolved into a spiral, and is equivalent to
nuclear spirals seen in the models built with the Eulerian scheme. In 
Maciejewski (2004b, hereafter M04b), I showed that nuclear ring
and spiral are two modes of wave propagation in centres of galaxies.
Here I focus on nuclear spirals, since nuclear rings have been already
studied in detail.

\section{Gas dynamics in nuclear spirals}
Generation of nuclear spirals can be well explained within the framework of
gas response to a fixed rotating stellar potential (Englmaier \& Shlosman 
2000). Gravity torques generate waves in gas that for any power-law rotation 
curve can propagate between the centre of the galaxy and the single Inner 
Lindblad Resonance present in that case (Maciejewski 2004a). The waves in this
region give rise to a spiral morphology in gas with the pitch angle 
proportional to the velocity dispersion in gas. Centres of non-axisymmetric 
galaxies act as resonant cavities, and generation of waves is inevitable there.

\begin{figure}
 \includegraphics[width=.49\linewidth]{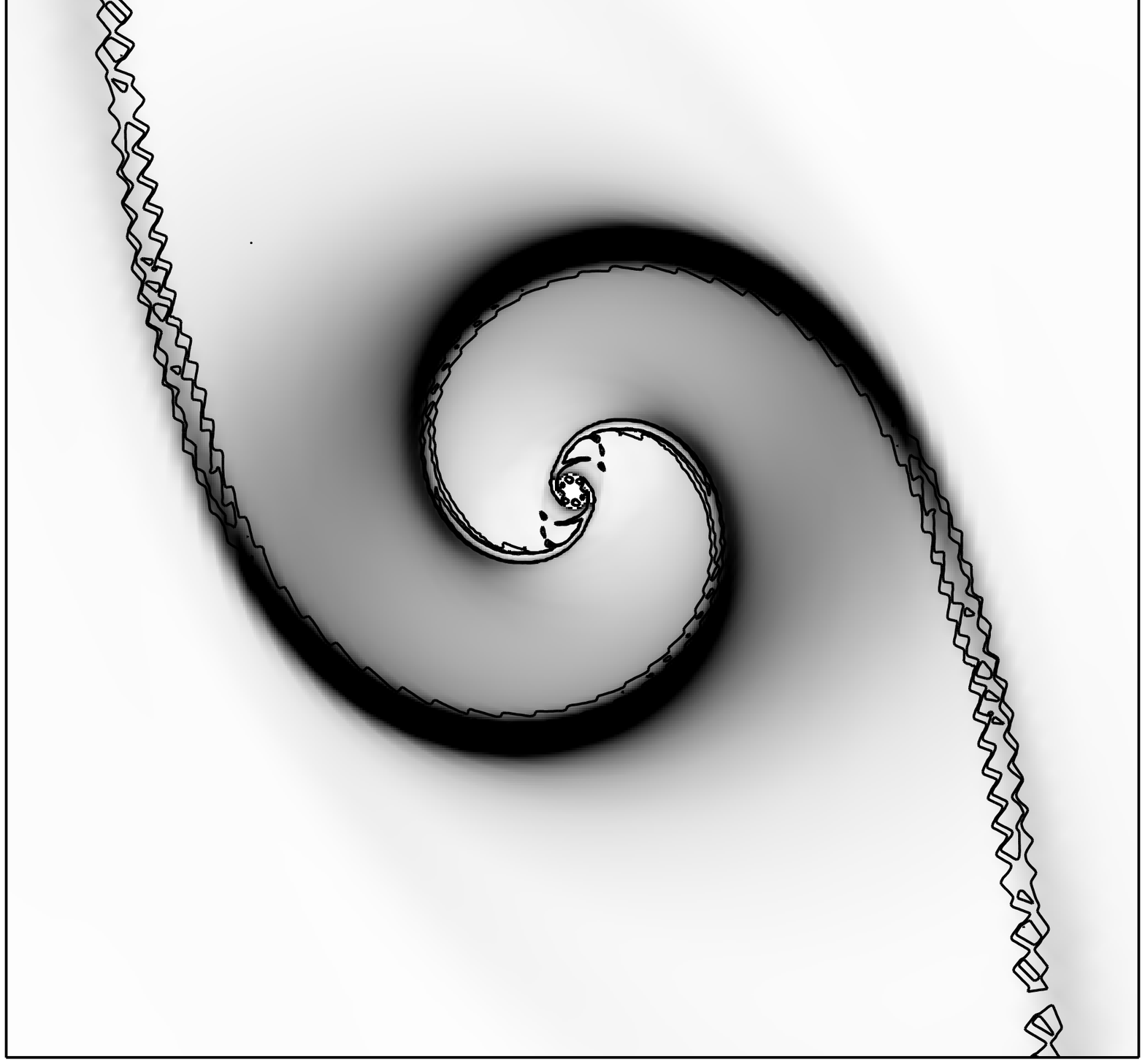}
 \includegraphics[width=.49\linewidth]{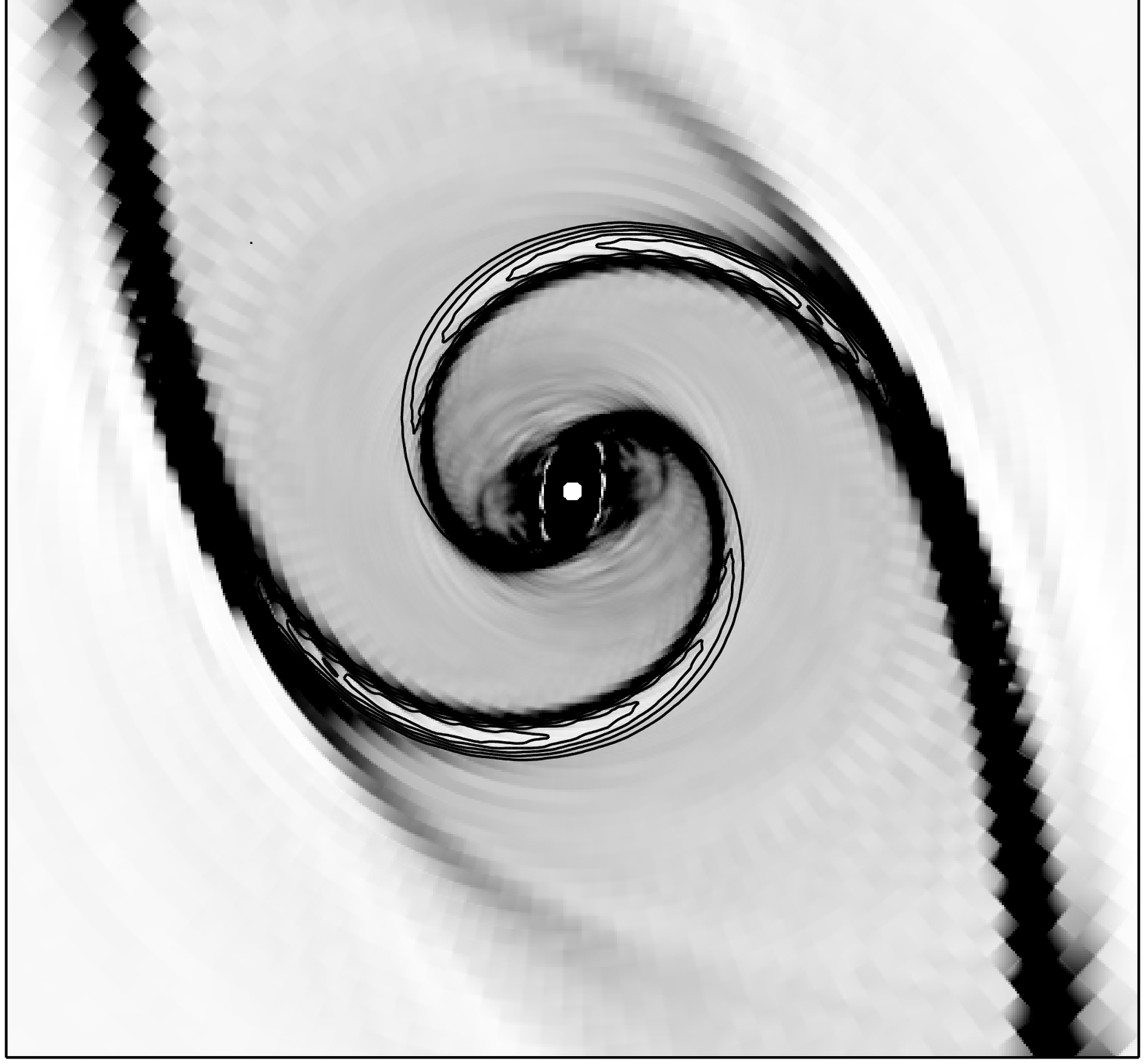}
 \vspace{-3cm}
  \caption{{\it Left:} A snapshot of gas density (greyscale) and div$^2 v$
(contours; only for div$v < 0$) in Model 8S20r of gas flow in a barred galaxy 
from M04b. Darker shading indicates higher density. 
{\it Right:} For the same model, $S^2$ (shear) is shown in greyscale, with the 
highest gas density overplotted in contours (unlike in the left panel). Darker 
shading indicates larger shear. The snapshot was taken at 0.5 Gyr, after the 
flow has stabilized. The bar is 6 kpc long and is vertical on the plots. The
box is 2 kpc long.}
\end{figure}

In M04b, I studied nuclear spirals with grid-based models constructed within 
the Eulerian scheme. I showed that models for small departures from axial 
symmetry are in excellent agreement with the analytical solution, which
provided an anchor for considering strongly barred models. Gas density and the 
location of shocks in nuclear spirals was given in M04b. Shocks were traced 
by large negative values of the divergence of gas velocity field. Divergence 
is a measure of expansion of the fluid, and its highly negative values should 
indicate gas compression in shocks. Here, I add one more characteristic of the
flow: its shear. The measure of shear in the plane of the galactic disc should 
take a form $\partial v_x/\partial y + \partial v_y/\partial x$, but this form 
is not invariant under rotation of the coordinate system. However, shear can 
still be characterized by a single invariant value for a two-dimensional flow 
in a plane. This value can be derived from the stress tensor 
$A_{ij}=\partial v_i/\partial x_j$, after extracting its symmetric part,
and leaving out the asymmetric one, responsible for the curl of the velocity 
field. Furthermore, the trace has to be separated out, since it solely 
contains terms corresponding to the expansion of the fluid, i.e. the
divergence of the velocity field. As a result, one is left with the tensor 
$D_{ij} = (A_{ij}+A_{ji})/2 - A_{kk} \delta_{ij}/2$, which describes the 
shear of the flow only. This is a traceless, symmetric tensor. The measure
of its magnitude can be obtained from its eigenvalues. In two dimensions,
the two eigenvalues of a traceless tensor have opposite signs, but the same 
magnitude $S$, which describes the amplitude of shear, and is expressed by
\[
S^2 = ( \frac{\partial v_x}{\partial y} + \frac{\partial v_y}{\partial x} )^2 +
  ( \frac{\partial v_x}{\partial x} - \frac{\partial v_y}{\partial y} )^2 .
\]

Here I calculated the shear of gas velocity field in Model 8S20r from 
M04b, where the nuclear spiral is present. Fig.1 displays
the gas density, square of divergence of the velocity field, div$^2 v$, as
the shock indicator, and $S^2$ as the indicator of the shear. Following the 
flow in Fig.1 from outside in, one can notice two almost straight lanes, with
density slightly above average, and strong shock and shear. These lanes,
almost vertical in Fig.1, mark the principal shocks in the bar that
often manifest themselves as a pair of straight dust lanes. As expected, the
highest density in gas is located downstream from the shock (gas rotates 
counterclockwise in Fig.1), and high shear extends downstream from the shock 
as well. When the flow is followed inwards, the principal straight shock
takes a spiral morphology, which I call the nuclear spiral (see also Englmaier
\& Shlosman 2000, Maciejewski et al. 2002, M04b). The divergence of gas 
velocity field remains highly negative, which 
indicates that the nuclear spiral is a shock in gas. The highest density 
in gas is still downstream from the shock, as expected. 

On the other hand, the distribution of shear in gas in the region of the
nuclear spiral is more complicated and very interesting. As the principal 
shock curves inwards and turns into the nuclear spiral, at azimuthal angles
somewhere half-way between the minor and the major axis of the bar, the thick 
dark lane marking large shear in the right-hand panel of Fig.1 splits. One
branch is exactly co-spatial with the shock -- this indicates large shear in 
the nuclear spiral shock. The other branch curves less than the spiral, splits
again, and disappears around the major axis
of the bar. Between these two branches of high shear, there is a pocket of
the lowest shear in the region, immediately downstream from the nuclear spiral
shock, and exactly co-spatial with the post-shock gas of highest density.
Thus the high-density post-shock gas that forms the gaseous nuclear spiral
experiences very little shear, contrary to the post-shock gas in the straight 
shock in the bar. This difference becomes clear once the region presented in 
Fig.1 is divided into elements, and for these elements the shear, quantified
by $S^2$, is plotted against overdensity, i.e. the ratio of the measured 
density to the initial one in the model, as in Fig.2. The points in the plot 
group 
in two perpendicular stripes: the vertical one, with overdensity about 10 and 
high shear, and the horizontal one, with low shear, below 0.5, but highest 
overdensities, reaching 100. It can be shown that the vertical stripe 
contains gas emerging from the straight shock in the bar, while the
post-shock gas in the nuclear spiral groups in the horizontal stripe.

\begin{figure}
\centering
 \includegraphics[width=.35\linewidth]{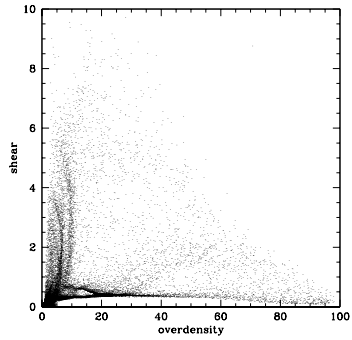}
  \caption{Shear as a function of overdensity for $512^2$ elements of the 
model from Fig.1.}
\end{figure}

\section{Conditions for star formation}
Presence of shocks and shear influences star formation: shocks 
trigger the collapse of molecular clouds, while shear may destroy the clouds. 
Thus one should expect enhanced star formation in the presence of shocks, but 
this may be inhibited by strong shear. In Model 8S20r, analyzed above, gas
orbiting around the galaxy centre passes through the nuclear spiral shock 
roughly every 10 Myr. Immediately downstream from the shock, it is compressed 
in gaseous spiral arms, whose density is a few times higher than the inter-arm
density, and where shear is negligible. These are excellent conditions for star
formation. Therefore in nuclear spirals star formation should occur on much 
higher rate than in the case when the inflowing gas gathers in an amorphous 
blob, which lacks shocks and post-shock high-density low-shear gas. Note 
however that star formation should not be confined to the gaseous spiral arms.
Its timescale is of the order of crossing time between the arms (10 Myr),
and one should expect that only the first step of star formation, the formation
of self-gravitating cores, occurs in the arms. However, young stars should 
appear at any azimuthal angle. Therefore, although star formation is
triggered by the spiral shock, it is distributed in a disc or wide ring.

The dynamics of nuclear spirals presented here is determined only by gas
response to an external gravitational potential. This process is well 
understood and its workings can be studied in detail. Obviously, self-gravity 
in gas, and feedback from star formation will make the picture presented here 
more complicated. However, once these processes are included, results become
uncertain, as 
star formation is poorly understood. Since in nuclear spirals compression of 
gas is induced by the spiral shock, the role of self-gravity in gas may not
be dominant there. Also the feedback from star formation is likely to be
weakened by the fact that most of the stars are not formed in the dense 
gaseous nuclear spiral.

\section{Conclusions}
I presented here the mechanism of central mass accumulation in nuclear spirals.
When a nuclear spiral is present in the centre of a galaxy, gas passes through
the nuclear spiral shock every 10 Myr or so, with no shear in the post-shock
compressed gas. This reoccurring condition for star formation makes something
like a cycle engine that efficiently converts gas into stars in the galaxy 
centre. Stars are also formed in nuclear rings, but the distribution of 
material
accumulated in nuclear spirals is more bulge-like, since the spiral spans over 
a considerable radial range. Star formation rate in nuclear spiral is likely to
be higher than in nuclear ring, since the nuclear spiral shock accompanies the 
gaseous spiral throughout its extent, while the nuclear ring damps the shock 
(see M04b).
Nuclear spirals might have played a crucial role in the early Universe, since
they form in gas with high velocity dispersion, likely present in newly formed 
galaxies. Ann (2005) also showed that nuclear spirals form preferably in
less massive galaxies. Like nuclear rings, nuclear spirals
may be responsible for the formation of pseudobulges, or, as it was pointed 
out during this meeting, disky bulges, since we still do not know how to
lift gas or stars from the galactic plane.


\begin{thebibliography}{}

\bibitem[ann05]{ann05}
     {Ann, H. B.} 2005,
     \textit{J. Korean Astronomical Soc.} 38, 121

\bibitem[ant05]{ant05}
     {Ann, H. B. \& Thakur, P.} 2005,
     \textit{ApJ} 620, 197

\bibitem[dmt07]{mdt07}
     {di Matteo, P., Combes, F., Melchior, A.-L. \& Semelin, B.} 2007,
     \textit{A\&A} 468, 61

\bibitem[eng00]{eng00}
     {Englmaier, P. \& Shlosman I.} 2000,
     \textit{ApJ} 528, 677

\bibitem[kor04]{kor04}
     {Kormendy, J. \& Kennicutt, R.C.} 2004,
     \textit{ARAA} 42, 603

\bibitem[mac00]{mac00}
     {Maciejewski, W.} 2000,
     in: F. Combes, G.A. Mamon \& V. Charmandaris (eds.),
     \textit{From the Early Universe to the Present},
     Proc. XVth IAP Meeting Dynamics of Galaxies (San Francisco: ASP), p.\ 63

\bibitem[mac4a]{mac4a}
     {Maciejewski, W.} 2004a,
     \textit{MNRAS} 354, 883

\bibitem[mac4b]{mac4b}
     {Maciejewski, W.} 2004b,
     \textit{MNRAS} 354, 892 (WM04b)

\bibitem[mac02]{mac02}
     {Maciejewski, W., Teuben, P.J., Sparke, L.S. \& Stone J.M.} 2002, 
     \textit{MNRAS} 329, 502

\bibitem[pat00]{pat00}
     {Patsis, P.A. \& Athanassoula, E.} 2000,
     \textit{A\&A} 358, 45

\bibitem[pin95]{pin95}
     {Piner, B.G., Stone, J.M. \& Teuben, P.J.} 1995,
     \textit{ApJ} 449, 508

\end{thebibliography}
\end{document}